\documentclass[twocolumn,aps,prb]{revtex4}
\usepackage{graphicx}
\usepackage{amssymb}
\usepackage{amsmath}
\usepackage{bm}
\usepackage{color}

\def\Journal #1,#2,#3,#4#5#6#7{#1 {\bf #2}, #3 (#4#5#6#7)}

\begin{document}

\title{Electron transmission through the stacking domain boundary in multilayers graphene}
\author{Nguyen N. T. Nam and Mikito Koshino}
\affiliation{Department of Physics, Tohoku University, Sendai 980-8578, Japan}
\date{\today}

\begin{abstract}
We present a theoretical study on the electron transmission through the AB-BA stacking boundary 
in multilayer graphenes. Using the tight-binding model and the transfer matrix method, we calculate the electron transmission probability through the boundary as a function of electron Fermi energy
in multilayers from bilayer to five-layer. We find that the transmission is strongly suppressed particularly near the band touching point, suggesting that
the electronic conductivity in general multilayer graphenes is significantly interfered by stacking fault.
The conductivity suppression by stacking fault is the strongest in the bilayer graphene, 
while it is gradually relaxed as increasing the number of layers.
At a large carrier density, we observe an even-odd effect
where the transmission is relatively lower in trilayer and five-layer than in bilayer and four-layer,
and this is related to the existence of a monolayer-like linear band in odd layers. 
For bilayer graphene, we also study the effect of the perpendicular electric field opening an energy gap,
and show that the band deformation enhances the electron transmission 
at a fixed carrier density. 
\end{abstract}

\maketitle

\section{Introduction}

After the successful isolation of individual graphene flake, graphene and its multilayers became one of the most intensively studied topics. In multilayer graphene, the weak van der Waals interlayer interaction allows domain structure consisting of different stacking regions. Recent experiments found that bilayer graphene samples commonly contain domain boundaries \cite{Alden2013, Lin2013}, which connect two distinct Bernal-stacking structures referred as AB and BA (Fig.\ \ref{fig:AB_boundary}).  In trilayer, similar stacking fault forms ABA-ABC stacking boundary which connects the Bernal stacked and rhombohedral stacked trilayer graphene \cite{Hattendorf2013}. This kind of domain structure is expected to rather common in graphene multilayers,
and it should influence the electric conductivity through the scattering of charge carriers.
The effect of the domain boundary on the electron transport was studied for graphene bilayer
\cite{Koshino2013,Jose2014}, and it was found that 
the transmission through the boundary is highly suppressed at the low carrier concentration.

In this paper,  we extend the previous analyses to general graphene multilayers from bilayer to five layer
and investigate the characteristic boundary effect on the electron transport in increasing the thickness.
Specifically we consider a multilayer system in
 Fig.\ \ref{fig:Boundary_trilayer} having AB-BA boundary on the bottom layer
while otherwise stacked in Bernal stacking order. 
We calculate the electron transmission probability through the stacking fault 
as a function of the Fermi energy using the tight-binding model and the transfer matrix method.
The result shows that the boundaries generally suppress the electron transmission in comparision with the perfect Bernal stacked system, while the reduction becomes gradually weaker as increasing the number of layers. We also notice that the transmission shows some odd-even effect at high electron density larger than $10^{13}$cm$^{-2}$, where the even-layer boundaries have better electron transmission than odd-layer boundaries.
We present the qualitative explanation of those characteristic properties 
of electron transmission through domain boundary by considering the local band structure inside the boundary region.

For Bernal stacked bilayer graphene, we can open an band gap 
between conduction band and valence bands 
by applying an external electric field perpendicular to the graphene sheet.
\cite{McCann_and_Falko_2006a,Guinea_et_al_2006a,Castro2007, Zhang2009, Lui2011,Xia2010,Wang2011,YanH2012,YanJ2012,Sugawara2011,Gong2012}. 
Here we also study the electron transmission in bilayer AB-BA boundary under the perpendicular 
electric field, and find that the band deformation induced by the electric field significantly 
enhances the electron transmission at a fixed carrier density.
\section{Formulation}

\subsection{Atomic structure}

		\begin{figure*}
		\begin{center}
			\leavevmode\includegraphics[width=0.7\hsize]{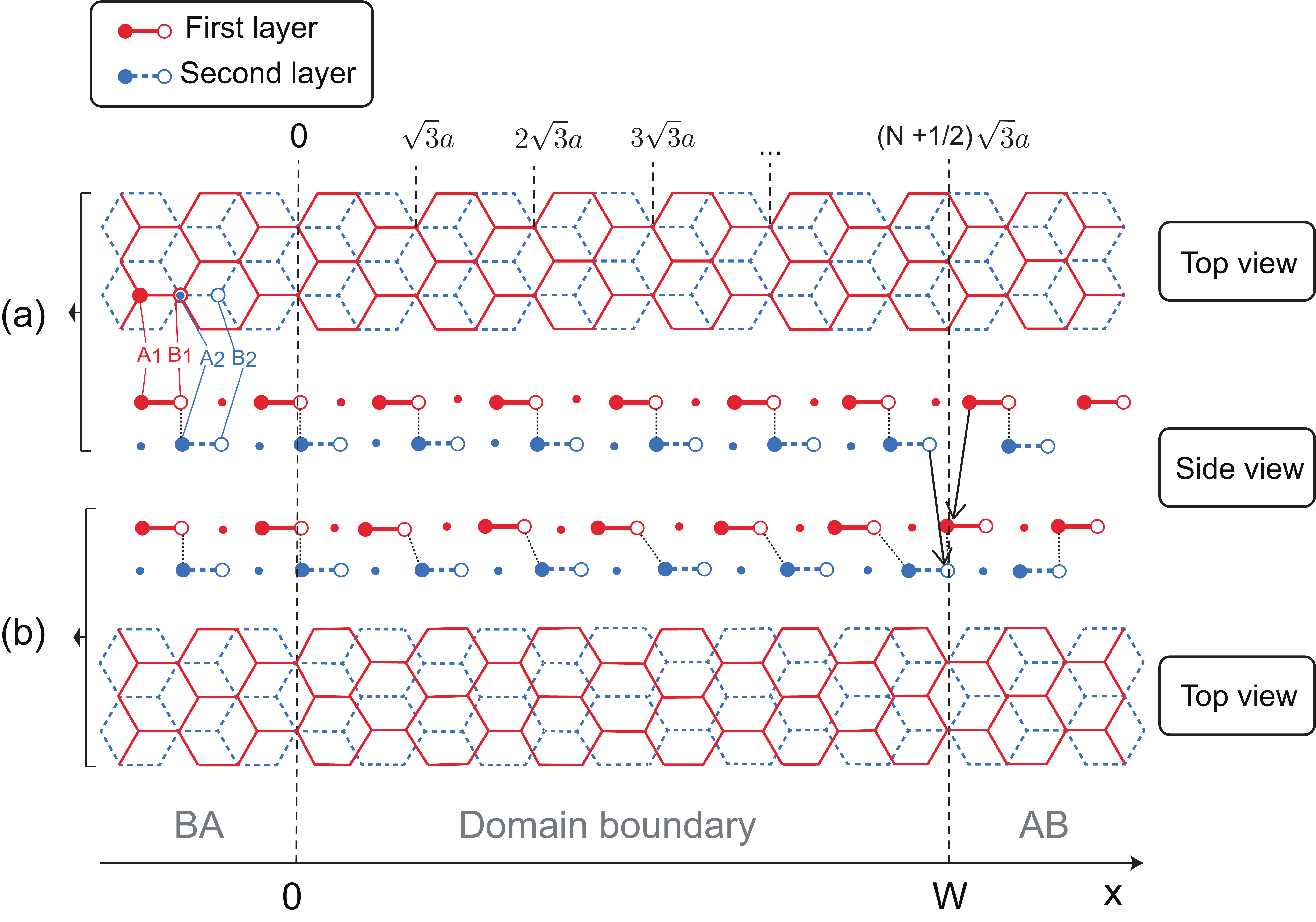}
		\end{center}
	\caption{Atomic structure of (a) regular BA bilayer graphene and (b) BA-AB domain boundary on bilayer graphene. \label{fig:AB_boundary}}
		\end{figure*}

				\begin{figure*}
		\begin{center}
			\leavevmode\includegraphics[width=0.6\hsize]{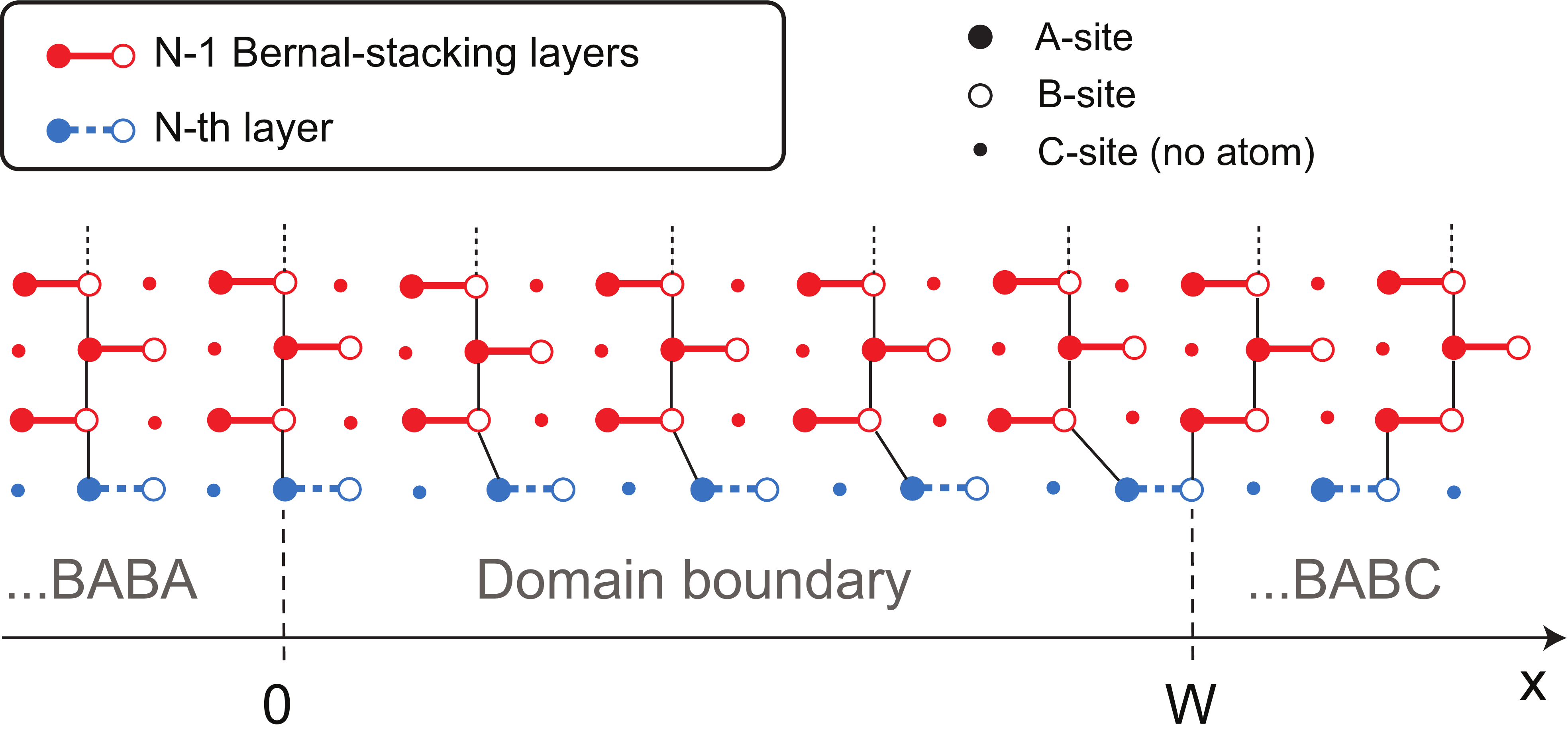}
		\end{center}
	\caption{Side view of N-layer domain boundary, which consists of $N-1$ Bernal stacking layers (red layers) 
	and one distorted layer at the bottom (blue layer). 
	\label{fig:Boundary_trilayer}}
		\end{figure*}		
		
Let us consider the AB-BA domain boundaries in bilayer graphene
as illustrated in Fig. \ref{fig:AB_boundary},
where the uniform BA stacking in the left region ($x<0$) is continuously deformed 
in the boundary region ($0<x<W$) to connect to the uniform AB stacking in the right region ($x>W$). 
We assume that the system is periodic in $y$-direction. 
The side view shows the atomic positions on a $xz$-plane, 
where filled circles and open circles indicate A and B atoms, respectively, 
while dots are the centers of hexagons denoted as C sites (no atoms). 
The domain boundary is created by shifting two graphene layers in the right region
with respect to each other, while fixing the left region.
Specifically, we start with the regular BA bilayer graphene [Fig \ref{fig:AB_boundary} (a)], 
and shift the top layer and  the bottom layer of the right region 
by $\pm a/(2\sqrt{3})$ (where $a = 0.246$ nm is the graphene's lattice constant), respectively, 
with the structure in the left region fixed.   
We assume that the atomic positions in the intermediate region
linearly interpolates between left and right.
The width of the intermediate region, $W$
is written in terms of an the number of unit cells $N$ as $W = \left(N +\frac{1}{2} \right)  \sqrt{3}a$.
At $N=20$, for instance, we have $W \approx 8.7$ nm,
while the realistic boundary width is about $10$nm \cite{Alden2013}. 
Here we focus on the boundary parallel to the zigzag direction which has 
the biggest effect on the transmission. \cite{Koshino2013}. 

The construction of the domain boundary in general $N$-layer graphene 
is similar to the bilayer case, where we assume 
only the bottom layer is expanded by $a/\sqrt{3}$ in $x$-axis 
while other $N-1$ layers are fixed to the  Bernal stacking without distorion
as illustrated in Fig.\ \ref{fig:Boundary_trilayer}.
The stacking structure can be specified by the sequence of sites along a vertical line;
here we see that $\cdots$BABA stack at $x=0$ 
is continuously deformed to $\cdots$BABC stack at $x=W$.


\subsection{Tight-binding model}
To model the motion of an electron in the system,
we use the tight-binding method. The Hamiltonian for tight-binding model is written as
\begin{equation}
H = -\sum_{i,j} t(\bm{R}_i -\bm{R}_j) |\bm{R}_i \rangle \langle \bm{R}_j| + \text{h.c.}
\end{equation}
where $\bm{R}_i$ is the atomic coordinate, $ |\bm{R}_i \rangle $ is the wavefunction at site $i$, and $t(\bm{R}_i -\bm{R}_j)$ is the transfer integral between atom $i$ and $j$. 
We adopt an approximation \cite{Moon2012, Nakanishi2001,Slater1954,Uryu2004,Trambly2010}
\begin{equation}
-t (\bm{d}) = V_{pp\pi} (d) \left[1 - \left(\frac{\bm{d} \cdot \bm{e}_z}{d}\right)^2 \right] +  V_{pp\sigma} (d)  \left(\frac{\bm{d} \cdot \bm{e}_z}{d}\right)^2
\end{equation}
\begin{align}
& V_{pp\pi} (d) = V_{pp\pi}^0  \exp \left(- \frac{d-a_0}{r_0} \right), \\
& V_{pp\sigma} (d)  =V_{pp\sigma}^0  \exp \left(- \frac{d-d_0}{r_0} \right)
\end{align}
where $\bm{d} = \bm{R}_i - \bm{R}_j$ is the distance between two atoms. $\bm{e}_z$ is the unit vector on $z$ axis. $V_{pp\pi}^0 \approx -2.7$eV is the transfer integrals between nearest-neighbor atoms of monolayer graphene which are located at distance $a_0 = a/\sqrt{3} \approx 0.142$nm. $V_{pp\sigma}^0 \approx 0.48 $eV is the transfer integral between two nearest-vertically aligned atoms. $d_0 \approx 0.335$nm is the interlayer spacing. The decay length $r_0$ of transfer integral is chosen at $0.184 a$. At $d > \sqrt{3}a$, the transfer integral is very small and can be neglected.

\subsection{Transmission probability}

\begin{figure}
		\begin{center}
		\leavevmode\includegraphics[width=0.9\hsize]{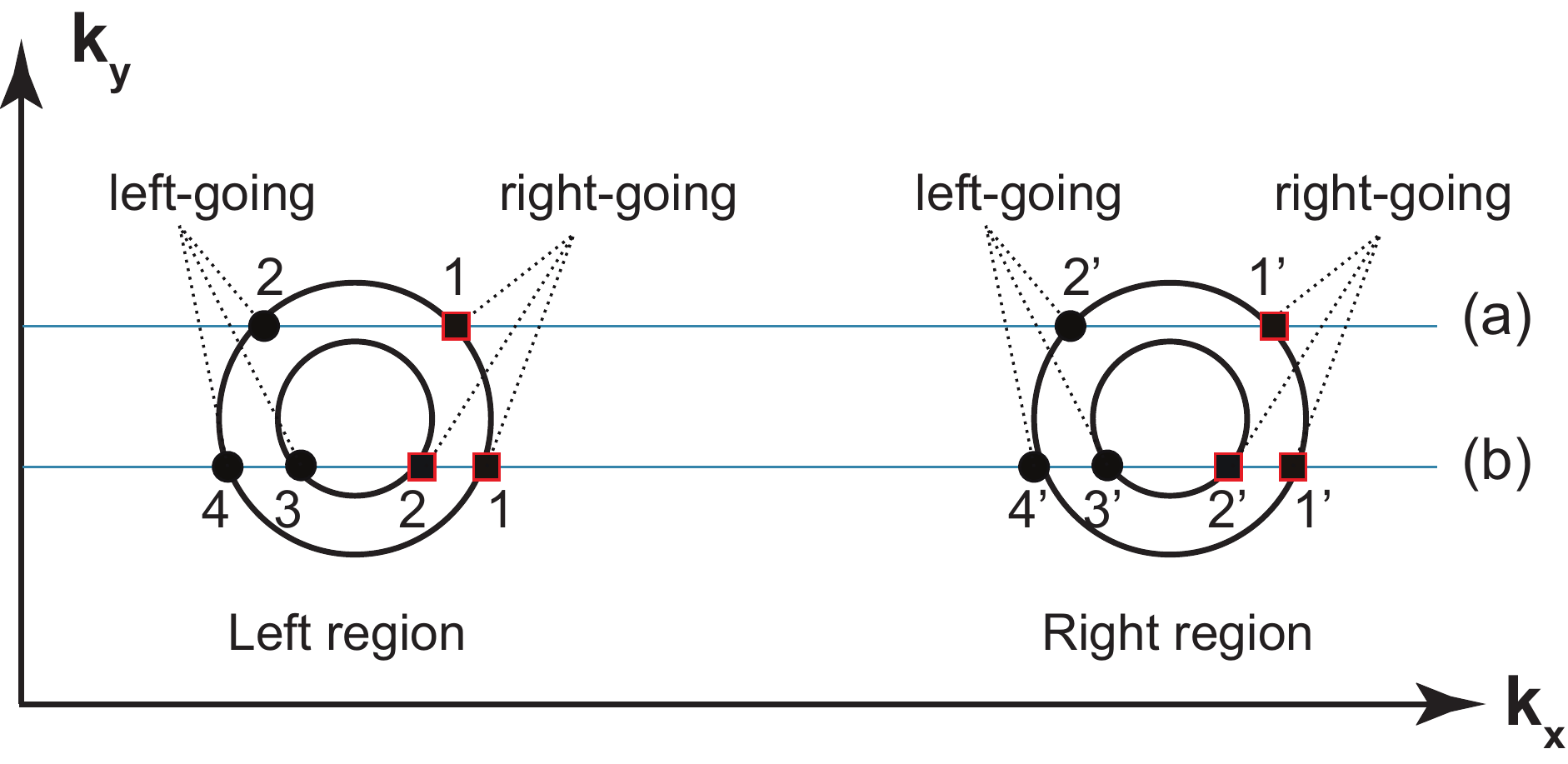}
		\end{center}
	\caption{Schematic view of Fermi surfaces of the left and right regions 
and traveling modes at a fixed energy and $k_y$.
\label{fig:Traveling_wave}}
	\end{figure}

Fig.\ \ref{fig:Traveling_wave}	 	
shows the schematic view of Fermi surface 
in the left region and the right region at a fixed energy $\varepsilon$.  
For $k_y$ falling inside the Fermi surface,
we have traveling modes with real $k_x$,
which can be classified as left-going and right-going modes
depending on the expectation value of velocity in $x$.
Since the system is translationally symmetric along $y$-direction,
the wavenumber $k_y$ is conserved, so that
an electronic state in the left region is only connected to one in the right region
on the same horizontal line.
Using the formulation in the Appendix,
we calculate the transmission coefficient $t_{\mu \nu} (k_y; \varepsilon)$ 
from incident channel $\nu$ in the left region to out-going channel $\mu$ in the right region.
The case (a) in Fig.\ \ref{fig:Traveling_wave}
shows a situation where we have a single channel per direction 
i.e, we have the right going mode 1 ($1'$) and the left going mode 2 ($2'$) 
in the left (right) region.
The electron transmission probability is then given by $|t_{1'1}|^2$.
In (b), we have multiple channels per direction; two right going modes 1, 2 ($1'$, $2'$) 
and two left going modes 3, 4 ($3'$, $4'$) in the left (right) region.
We then have multiple transmission probabilities
$|t_{1'1}|^2$, $|t_{2'1}|^2$, $|t_{1'2}|^2$, and  $|t_{2'2}|^2$.  

To characterize the total transmission for given boundary structure,
we calculate the Landauer's conductance 
\begin{align}
G(\varepsilon) 
&= \frac{e^2}{\pi \hbar} 
\sum_{k_y} \sum_{\mu \nu} |t_{\mu \nu} (k_y; \varepsilon)|^2
\nonumber\\
&\approx \frac{e^2}{\pi \hbar} 
 \int \frac{dk_y}{2\pi /L_y} \sum_{\mu \nu} |t_{\mu \nu} (k_y; \varepsilon)|^2,
\end{align}
where the indexes $\mu$ and $\nu$ run over all the traveling modes,
and we assumed the system is periodic in $y$-direction
with a sufficiently long period $L_y$.
The conductance $G$ is naturally proportional to $L_y$,
and the conductance per unit width, $G/L_y$, can be used as a quantity
which measures the transparency of the domain boundary
for the electronic transport.


\begin{figure}
		\begin{center}
			\leavevmode\includegraphics[width=1.\hsize]{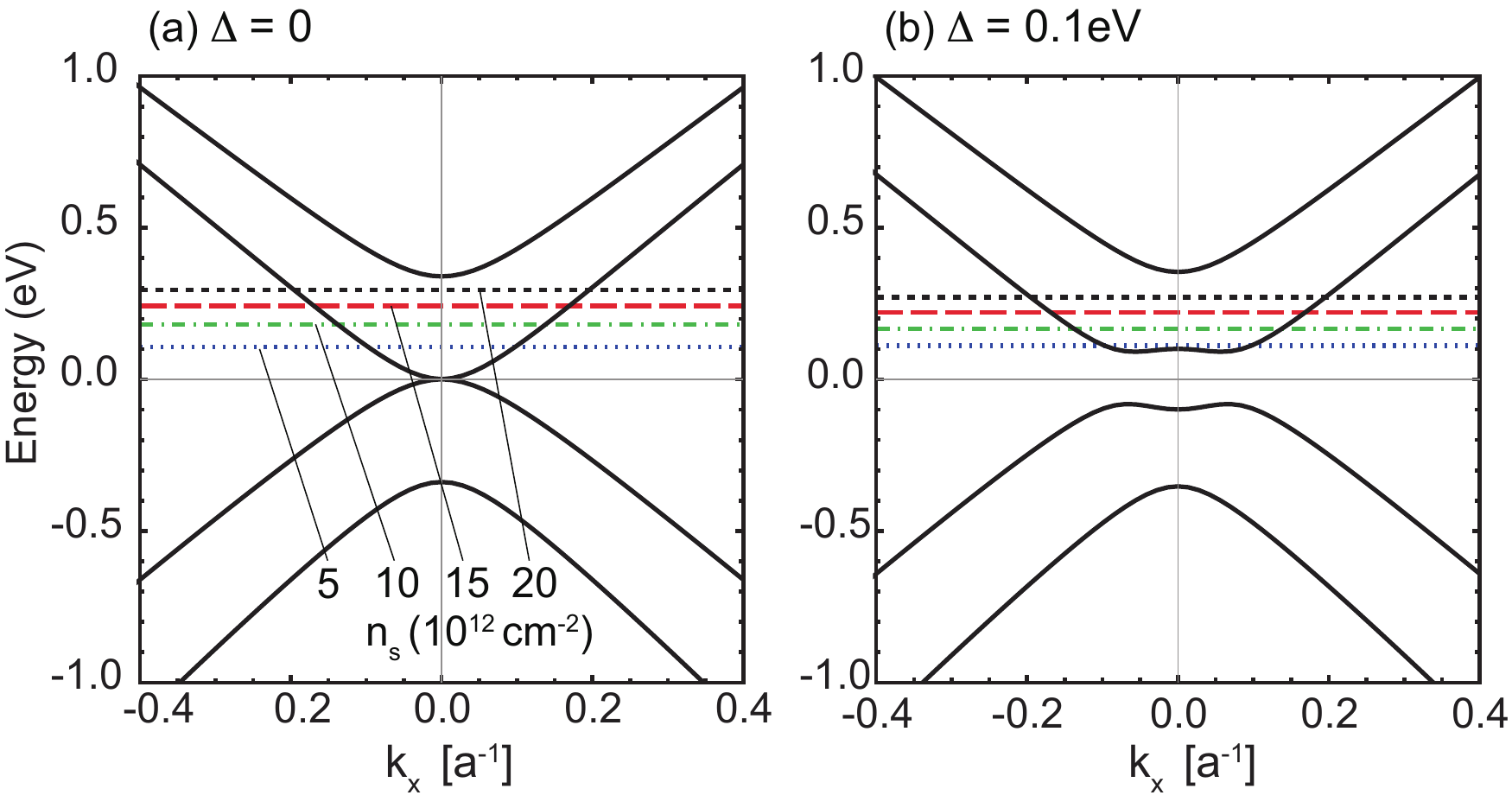} \\
		\end{center}
\caption{Energy band structures of bilayer graphene in the vicinity of K-point with
 the interlayer asymmetric potential (a) $\Delta=0$ and (b) $\Delta=0.1$ eV. 
 The dashed lines indicate the Fermi levels corresponding to several carrier densities $n_s$.
  \label{fig:Energy_band_2}}
\end{figure}

\begin{figure*}
	\begin{center}
		\leavevmode\includegraphics[width=0.8\hsize]{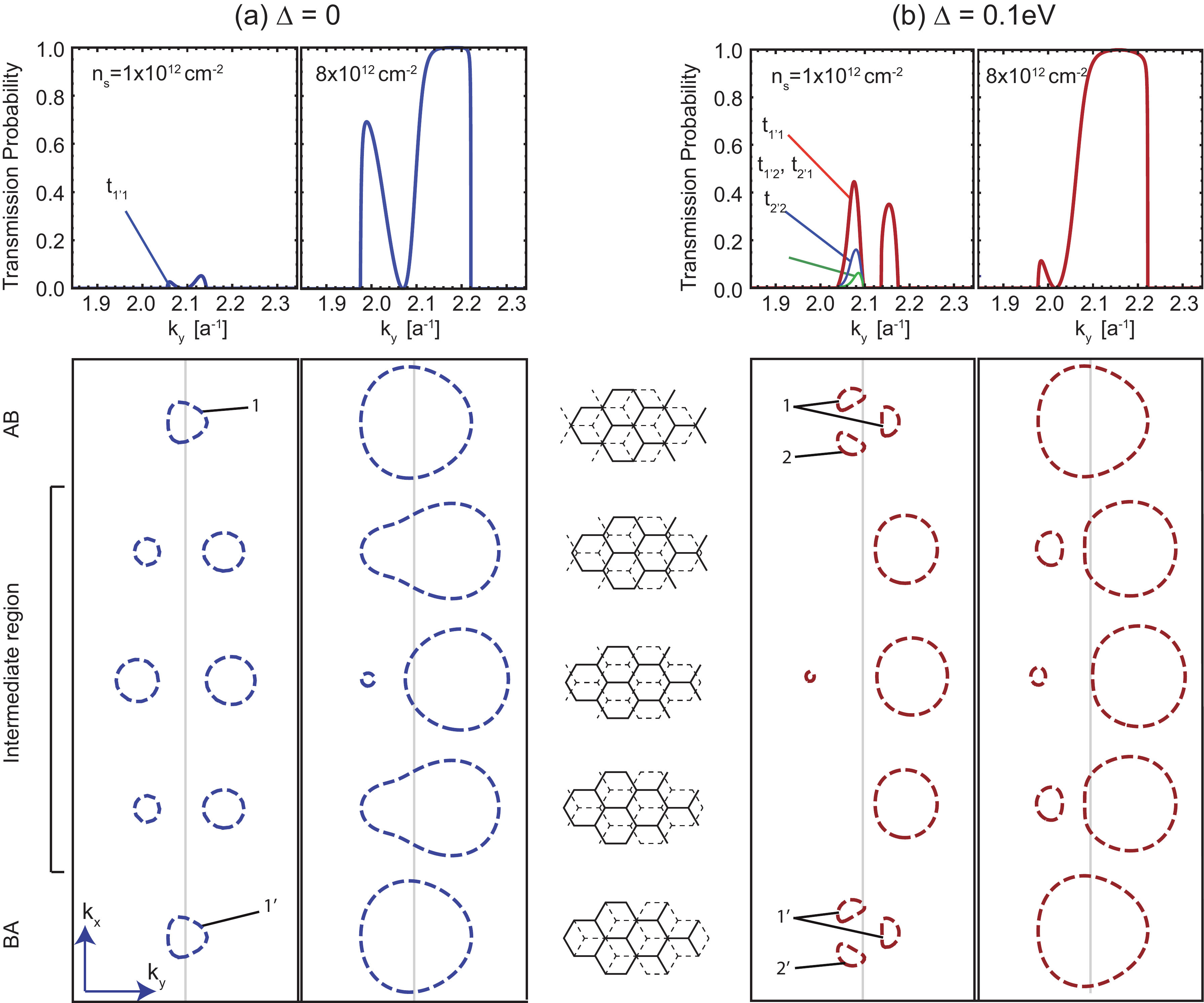}
	\end{center}
\caption{Electron transmission probability $|t_{\mu\nu}(k_y)|^2$ through AB-BA domain boundary 
($W=8.7$ nm) on the
bilayer graphene with the interlayer asymmetric potential (a) $\Delta=0$ and (b) $\Delta=0.1$ eV. 
The lower panels show the local Fermi surfaces and their corresponding local atomic structures.}	
\label{fig:TGLvsTGP}
\end{figure*}

\begin{figure}
	\begin{center}
		\leavevmode\includegraphics[width=0.9\hsize]{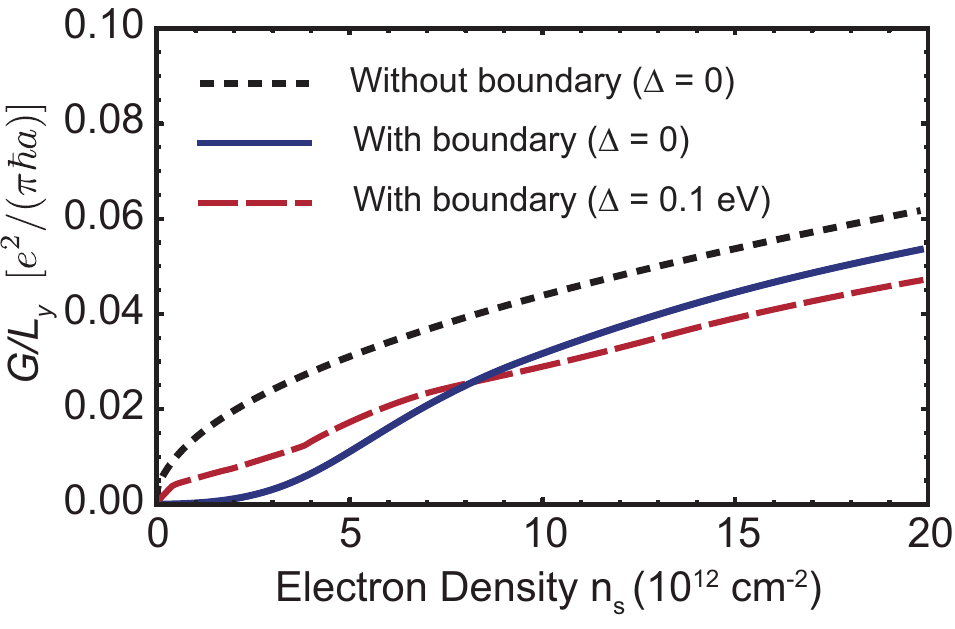}
	\end{center}
\caption{Total transmission $G/L_y$ as a function of the electron density
calculated for bilayer graphene of $\Delta=0$ with (blue, solid) and without (black dotted) AB-BA domain boundary
($W=8.7$ nm).
The red dashed curve is for the domain boundary
in presence of the interlayer asymmetric potential $\Delta=0.1$ eV. 
  }
\label{fig:Total_trans_3}
\end{figure}

\begin{figure*}
\begin{center}
\leavevmode\includegraphics[width=0.8\hsize]{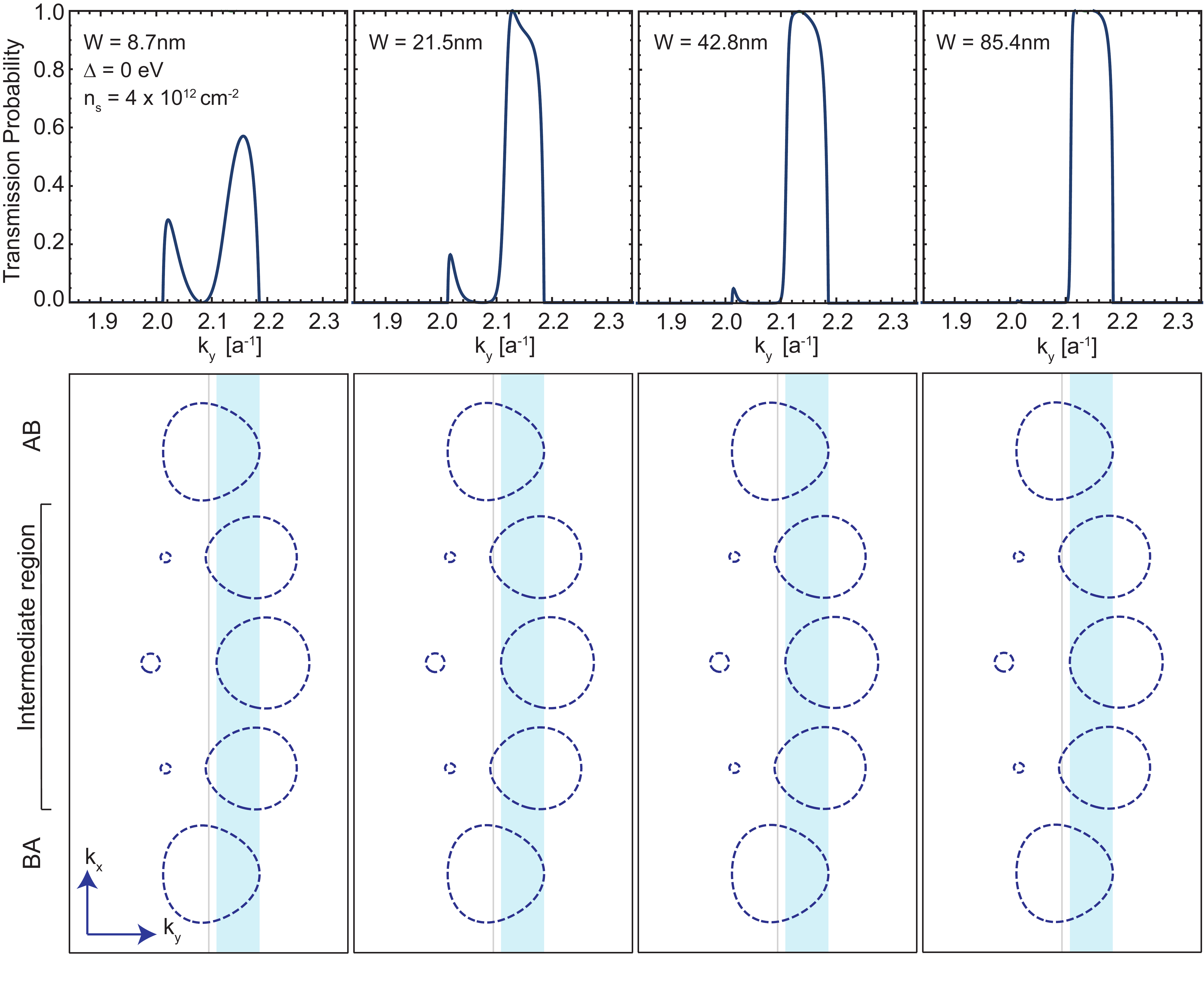}
\end{center}
\caption{(Top panels) Electron transmission probability $|t_{1'1}(k_y)|^2$ 
in the bilayer domain boundary with various widths $W$.
We fix $\Delta=0$ and $n_s = 4\times 10^{12}$cm$^{-12}$.
The blue shading shows the overlap region of the local Fermi sufaces (see text).
}
\label{fig:Width_changing_gapless6}
\end{figure*}

\section{Numerical result and discussion}

\subsection{Bilayer graphene and the effect of asymmetric electric potential}

We first consider the AB-BA domain boundary in bilayer graphene [Fig.\ \ref{fig:AB_boundary}]. 
The width of the boundary region is chosen as $N=20$, or $W = (N +1/2) a \sqrt{3} \approx 8.7$nm. 
Fig.\ \ref{fig:Energy_band_2} (a) shows the  
band structure of uniform AB-bilayer graphene,
where the energy origin $\varepsilon = 0$ is 
set to the band touching point. In the top panels of Fig.\ \ref{fig:TGLvsTGP} (a), we plot 
the transmission probability as a function of wave vector $k_y$
around the $K$-point at several Fermi energies.
Here we choose the energy range  $0 < \varepsilon_F < 0.15$eV,
where we have at most a single traveling channel 
in one-direction at each $k_y$,
where we have the only transmission probability $|t_{1'1}|^2$.

We observe that the electron transmission is very small near the bottom
of conduction band ($\varepsilon_F = 0.03$eV) where the electron waves 
mostly get reflected at the boundary, 
while the transmission probability sharply rises up in increasing the Fermi level. \cite{Koshino2013,Jose2014}
This behavior can also be seen in Fig.\ \ref{fig:Total_trans_3}
which plots the total transmission $G/L_y$ as a function of electron density. 
In Fig.\ \ref{fig:Total_trans_3}, we also compare the total transmission  
for the AB-BA domain boundary (blue)
with that for the uniform AB bilayer graphene without domain boundary (black).
We see that, that ratio of the former to the latter is almost zero
near the zero energy, while rapidly increases at higher energy levels.
This result suggests that the existence of the domain boundary dramatically reduces 
the conductivity of the bilayer graphene near the charge neutral point.

These characteristic features can be roughly explained from two different aspects.
First, we generally expect a bigger transmission
in a smoothly changing boundary
than in the sharply changing boundary.
The smoothness / sharpness  is characterized by
the ratio of the wave length of the incident wave $\lambda$
to the width of the boundary region $W$.
In the low Fermi energy ($\varepsilon_F = 0.03$eV),
the effective wavenumber measured from $K$ point 
is about $k \approx 0.04 a^{-1}$, 
and then the wavelength $\lambda = 2\pi/k \sim 40$ nm 
is actually larger than the width of
boundary region at $W \approx 8.7$ nm ($N=20$). 
So this case is regarded as a sharp boundary,
suggesting that we have a considerable reflection.
In increasing the Fermi level, the ratio $\lambda/W$
simply decreases, i.e., the boundary becomes relatively smoother
and gives a larger transmission.

Another important aspect is the wavenumber matching in local Fermi surface.\cite{Koshino2013}
In  Fig.\ \ref{fig:Boundary_trilayer},
we see that the stacking structure gradually changes from AB to BA
in increasing $x$, and the local structure at any particular $x$ approximates the uniform bilayer graphene
with a certain interlayer translation.
Then we can consider the local Fermi surface corresponding to the local atomic structure
as a function of $x$,
which is shown  in the lower panels of Fig.\ \ref{fig:TGLvsTGP}. 
At $\varepsilon_F = 0.03$ eV, for instance,
the Fermi surface in the intermediate region shifts along the $k_y$ direction,
leaving only a small overlap with initial Fermi surface
when projected to $k_y$ axis.
As a result, for a traveling wave of $k_y$ off the overlapping region
does not have corresponding traveling waves
in the intermediate region, then it does not reach the other side.
Meanwhile, at higher Fermi levels, the intermediate Fermi surface better overlaps 
with the initial one on the $k_y$ axis,
so that the boundary becomes more transparent.

The correspondence between the
transmission and the band structure 
is only approximate, but the agreement should become almost perfect
when $W$ becomes much larger than the typical wavelength.
To see this, we calculate the transmission probability 
at various widths of boundary region, $W$.
In Fig.\ \ref{fig:Width_changing_gapless6},
we show the transmission probability in the top panel,
and the local band structure in the bottom.
The blue shading indicates the region of $k_y$
which has traveling modes throughout the intermediate region.
For small boundary region W = 8.7 nm ($N=20$), 
we still have a finite transmission probability 
in the region outside of the blue shading region,
because there the wavelength $\lambda$ is actually larger than
$W$ as already argued, and thus the local Fermi surface is 
not well defined. 
Increasing $W$, the transmission outside the shaded region vanishes,
and at the same time the transmission probability in the shaded region
rapidly rises and reaches almost 1 at $W= 85.4$nm. 
This is because the wave is hardly reflected
in a smoothly-changing boundary
which is much longer than the electron wave length,
as long as a traveling mode exists throughout the 
intermediate region.

Next, we apply a perpendicular external electric field to the system by adding the 
asymmetric electrostatic potential $+\Delta$ and $-\Delta$ to the top and bottom layers.
As shown in Fig.\ \ref{fig:Energy_band_2} (b),
$\Delta$ actually opens up a band gap between the touching conduction and
valance band, and near the band edge, the energy dispersion exhibits a complex feature
where the Fermi surface split into several different parts due to the trigonal warping. \cite{McCann_and_Falko_2006a} 
In Fig.\ \ref{fig:TGLvsTGP} (b),
we compare the transmission probabilities in  gapped case ($\Delta=0$) and that in gapless case
($\Delta=0.1$ eV)  at the two different electron densities. 
At the lower density $n_s= 1.0 \times 10^{12}$cm$^{-2}$,
we have two right-going channels in the gapped case so that 
we plot multiple transmission probablities 
$|t_{1'1}|^2$, $|t_{2'1}|^2 = |t_{1'2}|^2$, and  $|t_{2'2}|^2$.  
We see that the transmission probability is much greater in the gapped case than in the gapless case. 
The total transmission calculation in Fig.\ \ref{fig:Total_trans_3}
 also shows that the suppression at low electron density observed in the gapless case
is considerably relaxed in the gapped case, suggesting that applying the external electric field definitely
increases the electron transmission near the charge neutral point. Here it should be noted that the total electron transmission can be changed purely by deforming the band structure, with the electron density fixed.
This feature can also be explained in terms of the local Fermi surface structure
shown in the lower panels Fig.\ \ref{fig:TGLvsTGP}.
At $n_s = 1.0 \times 10^{12}$ cm$^{-2}$, 
the Fermi surface spreads in a wider $k$-space region than in gapless case
due to the complex structure of band bottom.
This results in a shorter Fermi wavelength and 
also a better overlap with the intermediate Fermi surface,
and both of them contribute to a larger transmission.


\begin{figure}
\begin{center}
\leavevmode\includegraphics[width=1.0\hsize]{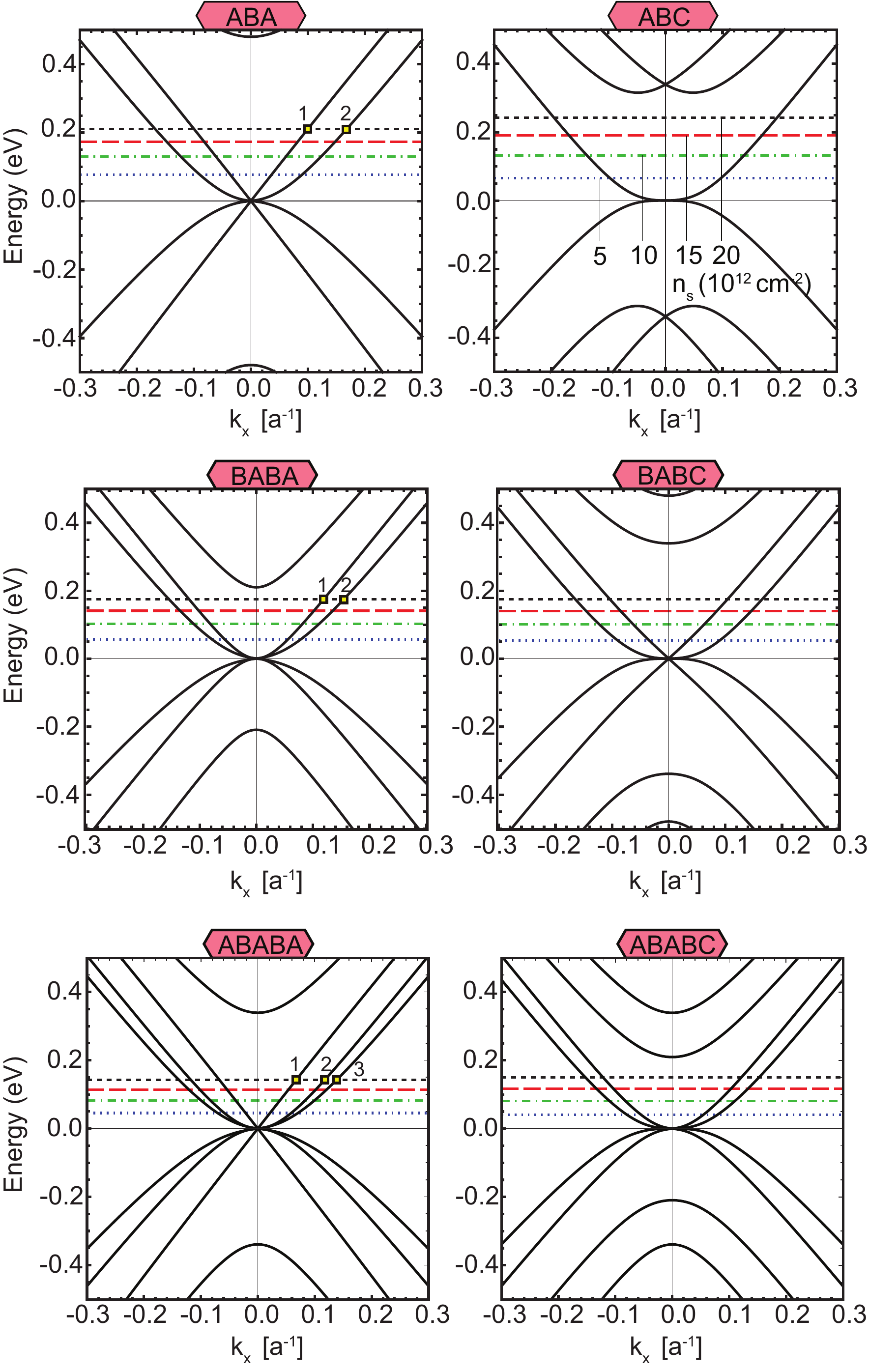}
\caption{Energy band structures in the vicinity of K-point of 
the left region (left panel) and right region (right panel) 
in trilayer (top), four-layer(middle) and five layer (bottom) domain boundary.
The dashed lines indicate the Fermi levels corresponding to several specific carrier densities $n_s$.	
	}
	\label{fig:Energy_bands_345}
\end{center}
\end{figure}

\begin{figure}
	\begin{center}
\leavevmode\includegraphics[width=1.0\hsize]{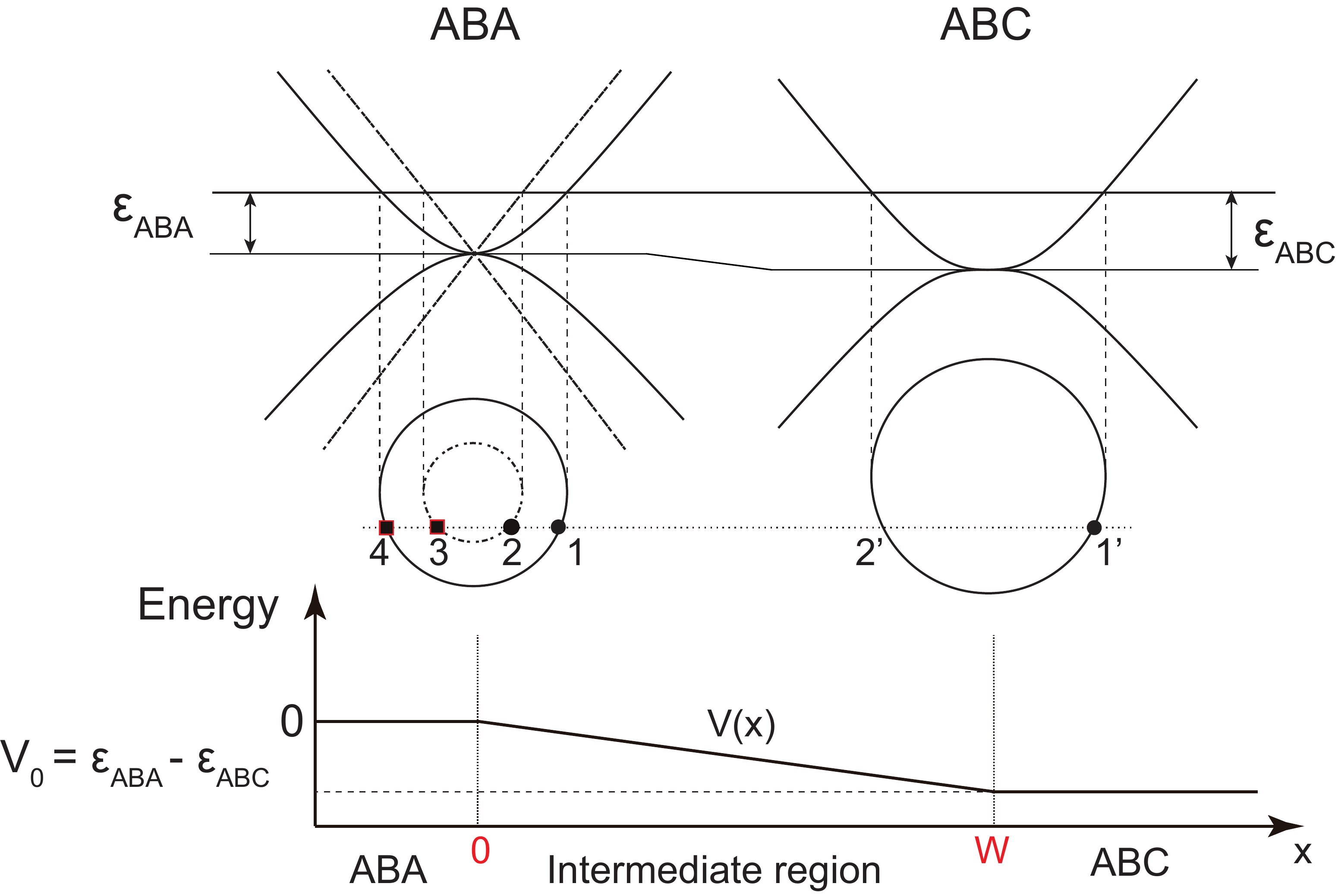}
	\end{center}
	\caption{Schematics of the electrostatic potential $V(x)$ to achieve the same electron density
	in ABA stack and ABC stack regions in trilayer graphene.
	\label{fig:Electric_field_trilayer}}
\end{figure}	

\begin{figure*}
\begin{center}
\leavevmode\includegraphics[width=0.8\hsize]{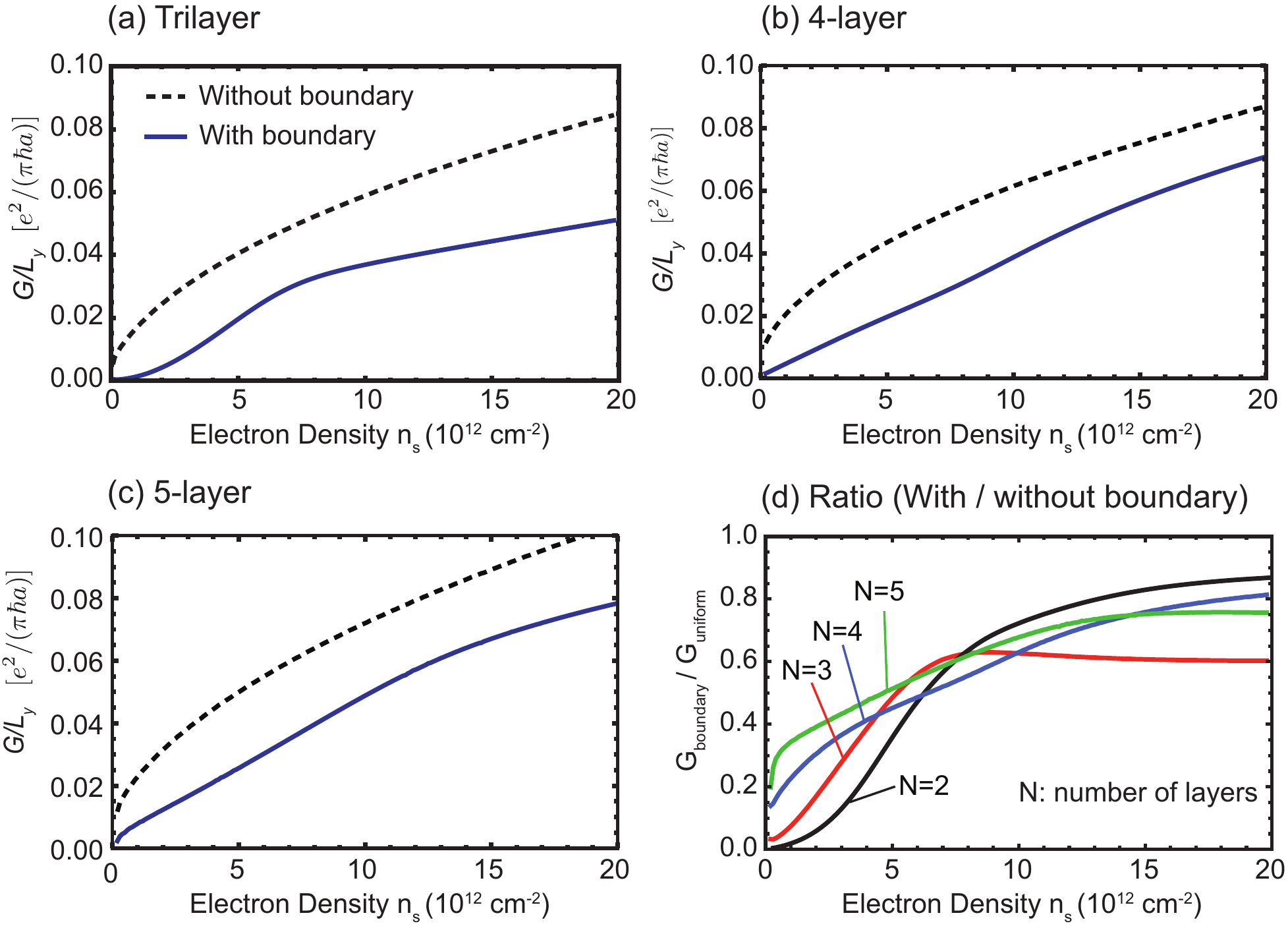}
\caption{
Total transmission $G/L_y$ as a function of the electron density
in (a) trilayer, (b) four-layer and (c) fivelayer,
where the black dotted and blue sold curves are for with and without domain boundary ($W=8.7$ nm), 
respectively.
(d) Ratio of $G$ with the boundary to that without the boundary 
from bilayer to five-layer.}
\label{fig:Totaltrans_multi}
\end{center}
\end{figure*}

\begin{figure}
\begin{center}
\leavevmode\includegraphics[width=0.7\hsize]{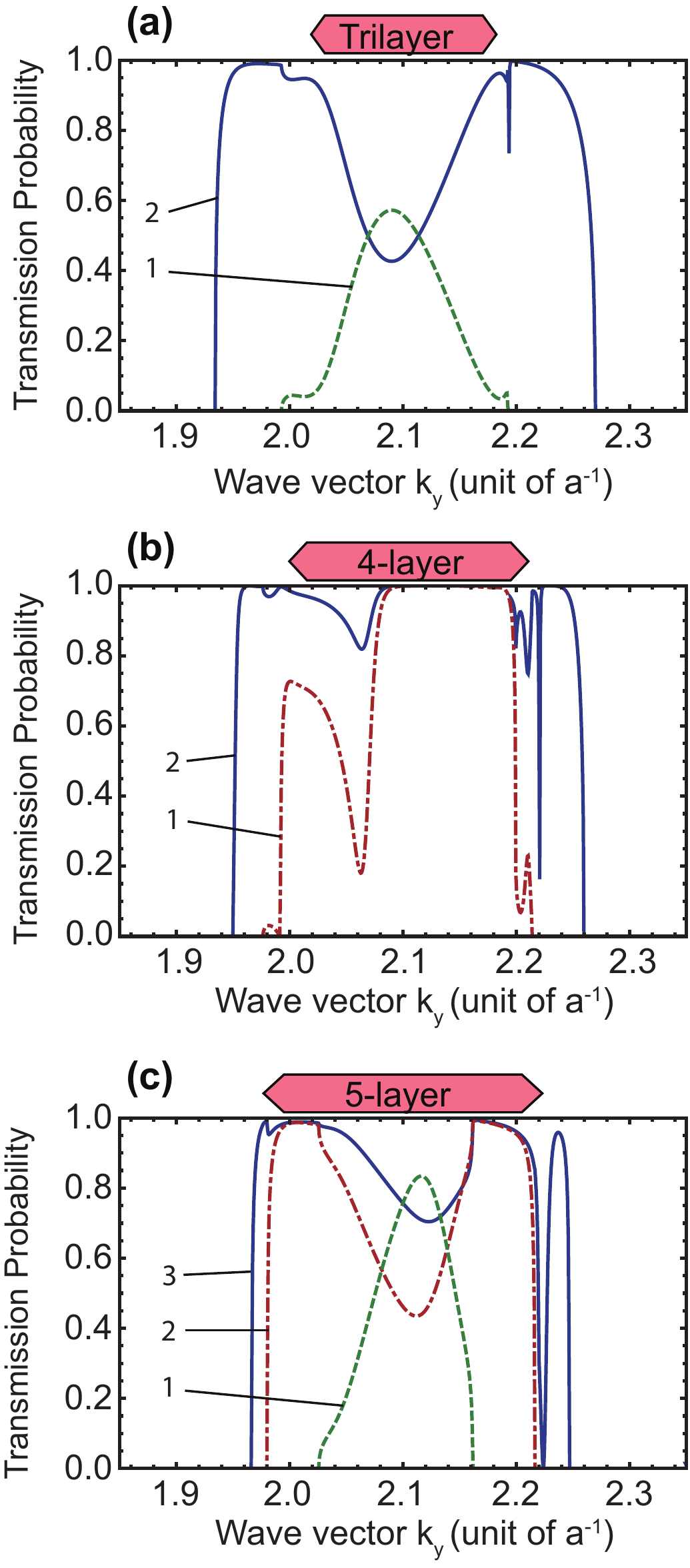}
	\caption{
Transmission probability $\sum_\mu |t_{\mu\nu}(k_y)|^2$ through the
domain boundary ($W=8.7$ nm)
from the incident channel $\nu$ in (a) trilayer, (b) four-layer and 
(c) five-layer at $n_s = 20 \times 10^{12}$ cm$^{-2}$.
The channel index $\nu = 1,2,\cdots$ corresponds to the numbers in Fig.\ \ref{fig:Energy_bands_345}.
}
\label{fig:Sum_trans}
\end{center}
\end{figure}

\subsection{Trilayer, four-layer and five-layer graphene}

In more than three-layer stack,
the left and right regions have generally different energy bands 
as shown in Fig.\ \ref{fig:Energy_bands_345}.
When we consider a realistic experimental situation
with a single gate electrode underneath the sample,
we have a homogeneous carrier density over the whole system, \cite{Nakanishi2010,Wolf2013}
where the left and right regions and share the same Fermi energy, 
while the origins of band energies are relatively shifted to achieve 
the same charge density in the both regions.
The situation is schematically illustrated in
Fig.\ \ref{fig:Electric_field_trilayer} for the trilayer case.
The shift of the band energy is given by the different electrostatic potentials
between the two regions, which automatically arises to satisfy the electrostatic equation
in the field-effect transistor geometry.
In the intermediate region, we simply assume the electrostatic potential $V(x)$ 
linearly changing in the intermediate region
to connect left and right as shown in the lower panel of Fig.\ \ref{fig:Electric_field_trilayer}, i.e.,
\begin{equation}
V(x) = \left\{ \begin{array}{ll}
0 \ \ \ \ \ \ \ (x < 0) \\
\frac{xV_0}{W} \ \ \ \ (0 \leq x \leq W) \\
V_0  \ \ \ \ \ (x > W) .
\end{array} \right., \label{V(x)}
\end{equation}
We calculate the transmission probability for the tight-binding Hamiltonian with $V(x)$ is included.

Figures \ref{fig:Totaltrans_multi} (a) (b) and  (c) plot
the total transmission $ G/L_y$ as a function of the electron density
for trilayer, four-layer and five-layer cases, respectively,
where the black and blue curves are for with and without domain boundary,
respectively. Figure\ \ref{fig:Totaltrans_multi} (d) shows 
the ratio of the transmission with the boundary to without the boundary.
The overall behavior in the trilayer boundary is pretty much 
similar to the bilayer case, i.e., the transmission is suppressed near the charge neutral point
and gradually rises at the high Fermi energy.
In adding more layer, however, we see that the reduction near $n_s=0$  becomes gradually weaker.
It is a natural consequence because the wave amplitude per single layer
becomes relatively smaller in a larger stack so that the transmission is less sensitive to the 
existence of the domain boundary on the surface layer.
When it comes to high density region, 
we observe some odd-even effect in Fig.\ \ref{fig:Totaltrans_multi} (d), 
where the boundaries with even number of layers show relatively better transmission 
than in the odd layers.
It is particularly conspicuous when comparing bilayer and trilayer,
where the transmission at $n=20 \times10^{12}$cm$^{-2}$ is 
almost 90\% in bilayer while it is only 60\% in trilayer.

The even-odd characteristics is closely related to
the left and right band structures presented in Fig.\ \ref{fig:Energy_bands_345}.
There the horizontal dashed lines indicate the Fermi energies 
corresponding to several carrier densities $n_s$.
In trilayer at $n_s  \leq 20 \times10^{12}$cm$^{-2}$, for example,
we have two right-going channels $\nu=1,2$ in the left region,
and a single right-going channel $\mu=1'$ in the right region.
We notice that the odd layer always has one more channel
in left region than in right region,
and this is due to the existence of the monolayer-like band 
in Bernal stacked odd layer graphene. \cite{Koshino2007,Koshino2008}
Fig.\ \ref{fig:Sum_trans} shows
the transmission probability $\sum_\mu |t_{\mu\nu}(k_y)|^2$
from the incident channel $\nu$ (summed over out-going channels $\mu$)
in trilayer, four-layer and five-layer at $n_s = 20 \times 10^{12}$ cm$^{-2}$.
In the odd-layers, 
the incident carriers from the linear band (green curve) are not very well transmitted,
and at the same time, the carriers from the other non-linear bands are considerably reflected 
only in the region of $k_y$ where a linear band state exists.
This suggests that the reflection matrix elements between the linear band 
and the other bands are  particularly strong, and it significantly interferes the total transmission.
The strong reflection in presence of the linear band is presumably related to the fact that 
a linear band state has a relatively large wave amplitude on the surface layer
compared to other bands. \cite{Koshino2008}
This fact explains why the transmission at the high carrier density is generally higher in even layer cases
where the linear band state is absent.


\section{Conclusion}
We studied the electron transmission properties in stacking domain boundary in
multilayer graphenes from bilayer to five-layer. 
We find the boundary significantly reduces the electron transmission at low Fermi energies,
 while the reduction becomes gradually weaker as increasing the number of layers. 
 We also find an odd-even effect at high electron density, 
 where the even-layer boundaries have better electron transmission than odd-layer boundaries.
For bilayer, we also demonstrated the conductance suppression at low energies
is significantly relaxed by applying a perpendicular external electric field.
We found that the transmission characteristics
is closely related the local electronic structure inside the boundary region. 
The agreement between the transmisssion and the overlap of the local band structure
is generally better for a wider boundary, and it becomes almost prefect when the boundary width
is much greater than the typical Fermi wavelength.

The study of the electron transmission through a single
domain boundary provides a very simple model 
to roughly understand the transmission properties of 
macroscopic multilayer graphenes and bulk graphite,
which are expected to have a number stacking faults and domains.
Our result indicates the electron transmission in a general graphitic system
 is significantly supressed in presence of domain boundaries. 
The present result also suggests that a boundary between two different stacking structures 
may provide a mechanism to control the electron current on multilayer graphene. 
In particular, the fact that the system being
almost insulator near bottom of conduction band and quickly becoming
transparent when increasing Fermi levels gives a possibility to be applied to switching devices. 

\appendix*

\section{Transfer matrix method}

\begin{figure}
\begin{center}
\leavevmode\includegraphics[width=0.9\hsize]{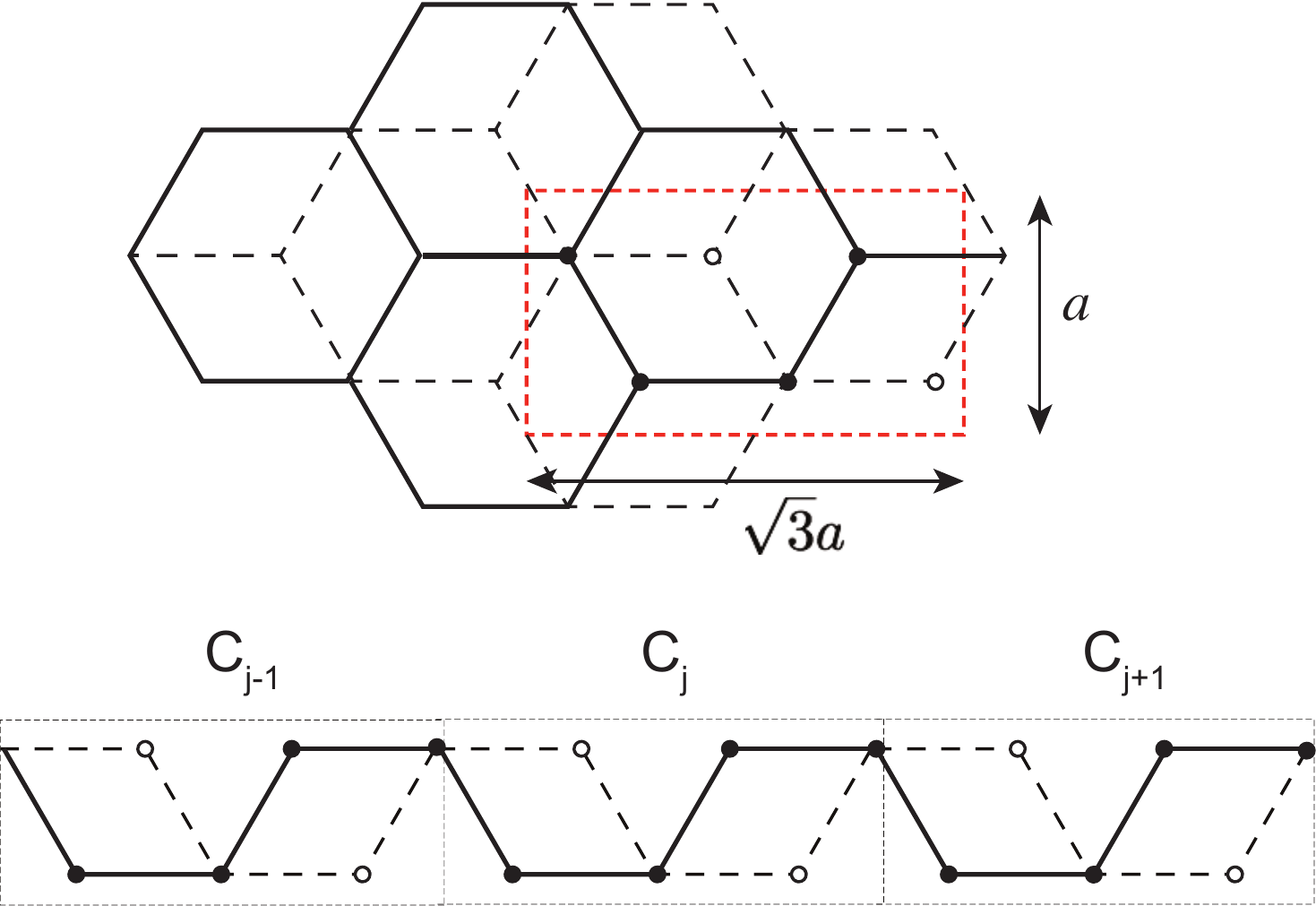}
\caption{
Unit cell of AB-bilayer graphene.
}
\label{fig:Unit_cell}
\end{center}
\end{figure}

In this section, we present the formula to calculate the 
transmission probability in the domain boundary. \cite{Ando1991}
We first consider the semi-infinite left and right regions separately,
and specify the traveling modes in each region.
Then we consider the intermediate region 
and calculate the transfer matrix connecting the traveling modes
in the left and right.

In uniform AB-stacked bilayer graphene, for example,
we define a unit cell as in Fig.\ \ref{fig:Unit_cell}, which consists of 
eight atoms. The Schr\"{o}dinger equation reads
				\begin{equation}
		(\varepsilon - \mathcal{H}_0) \bm{C}_j = - T^{\dagger}\bm{C}_{j-1} - T\bm{C}_{j+1},
				\end{equation}
	where
$\bm{C}_j$, $\bm{C}_{j-1}$, $\bm{C}_{j+1}$ are 8-component vectors 
consisting of 8 wavefunctions of 8 atoms in each unit cell, and
$\mathcal{H}_0$, $ T^{\dagger}$, $T$ are $8\times8$ matrices. $\mathcal{H}_0$ consists of transfer integrals between atoms inside each unit cells, while $T^{\dagger}$, $T$ consists of transfer integrals between atoms in $j$-th cell and atoms in $j-1$-th, $j+1$-th cell respectively. 
The equation is transformed as
\begin{equation}
 \lambda \left( \begin{array}{ll}
         \bm{C}_{j}\\
         \bm{C}_{j-1} 
         \end{array} \right) =  
         \left( \begin{array}{ll}
         -T^{-1} (\varepsilon - \mathcal{H}_0) & - T^{-1} T^{\dagger}\\
        \ \ \ \ \ \ \ \ \  1 & \ \ \ \ \ 0  
         \end{array} \right)\left( \begin{array}{ll}
         \bm{C}_{j}\\
         \bm{C}_{j-1} 
         \end{array} \right),
\end{equation}
which gives 16 eigenvalues of $\lambda$. 
These modes are classified as traveling modes 
if $|\lambda| = 1$, and 
as evanescent modes if $|\lambda| \neq 1$.
For traveling modes, we can define the real wave number $k_x$
by $\lambda = \exp(ik_x \sqrt{3}a)$,
and it is even classified to left-going or right-going modes
depending on the expectation value of the velocity along $x$-direction.
For evanescent modes, we categorize $|\lambda| > 1$ and $|\lambda| < 1$
as left-going and right-going modes, respectively,
for the sake of convenience.
The 16 eigenmodes are always composed of 8 right-going and 8 left-going modes. 
Let $\lambda^\pm_1$, $\lambda^\pm_2$, ..., $\lambda^\pm_8$
be the eigenvalues of the right and left going modes, respectively.

The wavefunction is written as
$\bm{C}_j = \lambda^j\Vec{u}$ with 8-component eigenvector $\Vec{u}$.
We define $\bm{u}_i^\pm$ as the eigenvectors corresponding to  $\lambda_i^\pm$.
We then define the $8\times8$ matrices
\begin{equation}
U^{\pm} = ( \bm{u}_1^{\pm}, \bm{u}_2^{\pm} \cdots  \bm{u}_8^{\pm}),
\end{equation}
\begin{equation}
\Lambda^{\pm} = \left(
 \begin{array}{ccccc}
   \lambda^{\pm}_1 \\
    & & \ddots & & \\
    & & & & \lambda^{\pm}_8
 \end{array}
\right).
\end{equation}
Now any left and right-going waves can be written as the superposition of the waves functions $\bm{u}_j^-$ or $\bm{u}_j^+$. For example, wave function at site 0
is written as. \begin{equation}
\bm{C}^{\pm}_0 = \sum_i^8 \bm{u}^{\pm}_i \alpha_i = U^{\pm} \bm{\alpha}^{\pm},
\end{equation}
where 
\begin{equation}
\bm{\alpha}^{\pm} = \left( \begin{array}{ll}
         \alpha^{\pm}_1 \\
         \vdots\\
         \alpha^{\pm}_8 
         \end{array} \right).
\end{equation}
Then the wave amplitude at $j$-th site becomes
\begin{equation}
\bm{C}^{\pm}_j = \sum_i^8 (\lambda^{\pm})^j \bm{u}_i^{\pm} \alpha_i^{\pm} = U^{\pm} (\Lambda^{\pm})^j \bm{\alpha}^{\pm},
\end{equation}
leading to the relation
\begin{equation}
\bm{C}^{\pm}_{j+1} = [ U^{\pm} (\Lambda^{\pm})^j (U^{\pm})^{-1}]\bm{C}^{\pm}_j \equiv {F}^{\pm}\bm{C}^{\pm}_j, \label{Bloch_theorem}
\end{equation}
which is similar to the Bloch theorem. 
The wavefunction at site $j$ can be generally written as summation of left-going wave and right-going wave.
		\begin{equation}
		\bm{C}_j = \bm{C}_j^+ + \bm{C}_j^-
		\end{equation}

By doing the similar process for BA bilayer graphene (the right region), we are also able to get the eigenvectors and eigenvalues. 
We define $U^{\pm}_L$, $\Lambda^{\pm}_L$, ${F}^{\pm}_L$
for the left region (AB-stack),
and $U^{\pm}_R$, $\Lambda^{\pm}_R$, ${F}^{\pm}_R$
for the right region (BA-stack).

\begin{figure}
		\begin{center}
			\leavevmode\includegraphics[width=0.9\hsize]{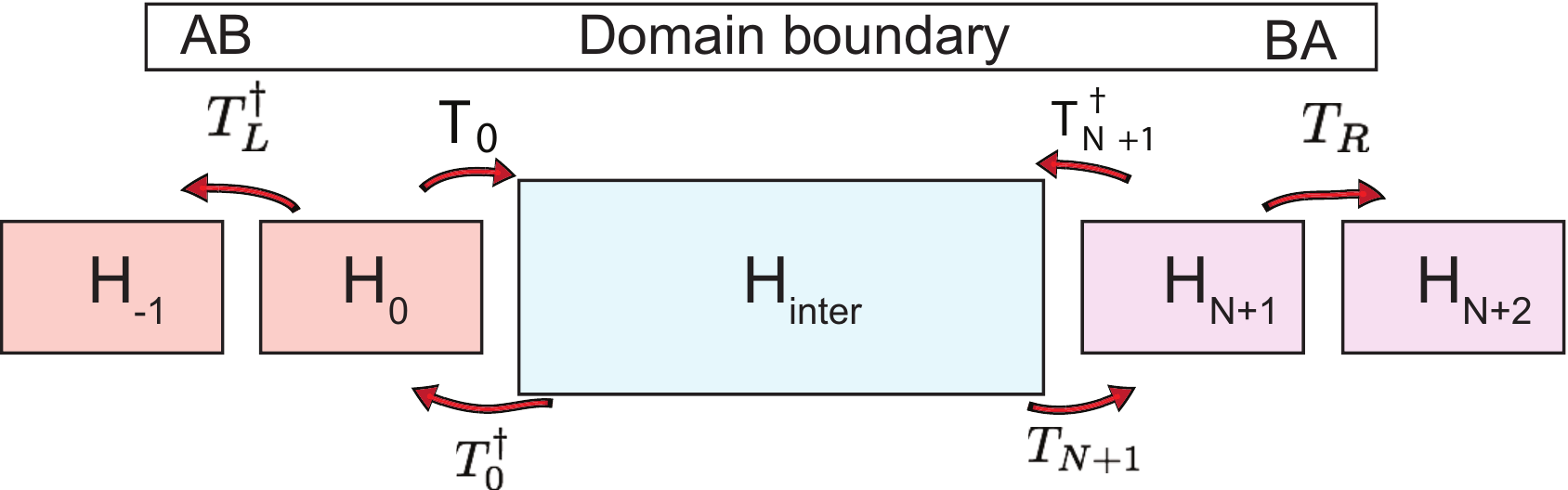}
		\end{center}
	\caption{Schematic view of the AB-BA domain boundary.\label{fig:The_system}}
		\end{figure}

Figure \ref{fig:The_system} schematically illustrates 
the Hamiltonian of the AB-BA domain boundary.
The tight-binding equations read
		\begin{equation}
	\left\{ \begin{array}{ll}
         (\varepsilon -\mathcal{H}_0) \bm{C}_0 = - T_0 \bm{C}_{\text{inter}} - T_L^{\dagger} \bm{C}_{-1}, \\
         (\varepsilon -\mathcal{H}_{\text{inter}}) \bm{C}_{\text{inter}} = - T_0^{\dagger} \bm{C}_{0} - T_{N+1} \bm{C}_{N+1}, \\
         (\varepsilon -\mathcal{H}_{N+1}) \bm{C}_{N+1} = - T^{\dagger}_{N+1} \bm{C}_{\text{inter}} - T_{R} \bm{C}_{N+2}. \\
            \end{array} \right. \label{Tight_binding_system}
		\end{equation}
Here 
$\bm{C}_{0}$, $\bm{C}_{-1}$, $\bm{C}_{N+1}$, $\bm{C}_{N+2}$ are
8-component vectors, $\bm{C}_{\text{inter}}$ is $n$-component vector,
with $n$ being the number of atoms in the intermediate region.
$\mathcal{H}_0$, $\mathcal{H}_{N+1}$ are $8 \times 8$ matrices, 
$\mathcal{H}_{\text{inter}}$ is a $n\times n$ matrix,
and $\mathcal{T}_0$, $\mathcal{T}_{N+1}$ are $n \times 8$ and 
$8 \times n$ matrices, respectively.

In the following, we consider a situation where 
the incident wave comes from the left region, and transmits to the right
or reflects back to the left.
From (\ref{Bloch_theorem}), we have 
\begin{equation}
\left\{ \begin{array}{ll} 
\bm{C}_{-1} = ({F}^+_L)^{-1}\bm{C}_0^+ +  ({F}^-_L)^{-1}\bm{C}_0^- \\
\qquad\quad = ({F}^+_L)^{-1} \bm{C}_0^+ + ({F}_L^-)^{-1} (\bm{C}_0 - \bm{C}_0^+), \\
\bm{C}_{N+2} = {F}^+_R \bm{C}^+_{N+1} ={F}^+_R \bm{C}_{N+1}.
  \end{array} \right. 
\end{equation}
Note that there are only right-going waves in the right region.
Then (\ref{Tight_binding_system}) becomes
		\begin{equation}
	\left\{ \begin{array}{ll}
         (\varepsilon -\mathcal{H}_0 + T_L^{\dagger}({F}^+_L)^{-1}) \bm{C}_0
        \\ \qquad  = - T_0 \bm{C}_{\text{inter}} - T_L^{\dagger} [({F}^+_L)^{-1}) -({F}_L^-)^{-1} ]\bm{C}_{0}^{+}, \\
         (\varepsilon -\mathcal{H}_{\text{inter}}) \bm{C}_{\text{inter}} = - T_0^{\dagger} \bm{C}_{0} - T_{N+1} \bm{C}_{N+1}, \\
         (\varepsilon -\mathcal{H}_{N+1} +  T_{R}{F}^+_R) \bm{C}_{N+1} = - T^{\dagger}_{N+1} \bm{C}_{\text{inter}}.  \\
            \end{array} \right. 
		\end{equation}	
	which leads to
		\begin{equation}
	\left( \begin{array}{ll}
         \bm{C}_0\\
       \bm{C}_{\text{inter}} \\
        \bm{C}_{N+1} 
         \end{array} \right) = 
        \frac{1}{E - \tilde{\mathcal{H}}}\left(   \begin{array}{ll}
        -T_L^{\dagger} [({F}^+_L)^{-1}) -({F}_L^-)^{-1} ]\bm{C}_{0}^{+}  \\
         \ \ \ \ \ \ \ \ \ \ \ \ \ 0 \\
        \ \ \ \ \ \ \ \ \ \ \ \ \  0
         \end{array}  \right) .
	\end{equation}	 
The Hamiltonian is $(n+16)\times(n+16)$ matrix
\begin{equation}
\tilde{\mathcal{H}} = \left( \begin{array}{ccccc}
\mathcal{H}_0 - T_L^{\dagger}({F}^+_L)^{-1}   & \ \ \ \ -T_0 & 0 \\
\\ 
-T^{\dagger}_0 & \ \ \ \  \mathcal{H}_{\text{inter}} & -T_{N+1} \\
\\
0 & \ \ \ \  -T_{N+1}^{\dagger} & \ \ \ \ \ \ \mathcal{H}_{N+1} -  T_{R}{F}^+_R
\end{array}  \right).  \label{Hamiltonian_system}
\end{equation} 	
	We define the Green function 
	\begin{equation}
	G =  \frac{1}{E - \tilde{\mathcal{H}}} = \left( \begin{array}{cccc}
	G_1 & \cdots & \\
	\vdots & \ddots & \vdots \\
	G_2 & \cdots
	\end{array} \right),
	\end{equation}
where $G_1$ and $G_2$ are $8\times 8$ matrices.
	
The wavefunction at site $j=0$ can be written as
\begin{equation}
\bm{C}_0 = \bm{C}_0^+ + \bm{C}_0^- = -G_2T_L^{\dagger} [({F}^+_L)^{-1} -({F}_L^-)^{-1} ]\bm{C}_{0}^{+}.
\end{equation}
Then we have
\begin{equation}
\bm{C}_0^- = \{-G_2T_L^{\dagger} [({F}^+_L)^{-1} -({F}_L^-)^{-1} ] - 1 \}\bm{C}_{0}^{+}.
\end{equation}
Similarly, the right-going waves at site $N+1$ is writte nas
\begin{equation}
\bm{C}_{N+1}^+ = \bm{C}_{N+1} = -G_1 T_L^{\dagger} [({F}^+_L)^{-1} -({F}_L^-)^{-1} ]\bm{C}_{0}^{+}.
\end{equation}
To calculate the transmission and reflection probability, we first write the left and right-going waves at site 0, N + 1 in terms of left-going waves and right-going waves
\begin{align}
\bm{C}_0^+ = U^+_L \bm{\alpha}_i, \\ 
\bm{C}_0^- = U^-_L \bm{\alpha}_r, \\
\bm{C}_{N+1}^+ = U^+_R \bm{\alpha}_t.
\end{align}
The transmission and reflection matrices are defined by
$\bm{\alpha}_t = T \bm{\alpha}_i$ and $\bm{\alpha}_r = R \bm{\alpha}_i$,
and they are given by
	\begin{align}
	& \bm{T} =   (U_R^+)^{-1} \left( -G_2.T_L^{\dagger} [({F}^+_L)^{-1} -({F}_L^-)^{-1} ]\right) U^+_L, \\
	& \bm{R} =  (U_L^-)^{-1} \left( -G_1.T_L^{\dagger} [({F}^+_L)^{-1} -({F}_L^-)^{-1} ] - 1 \right)U_L^+.
	\end{align}
The transmission coefficient for incident wave $\nu$ with velocity
$v_{\nu}$ and out-going channel $\mu$ with velocity $v_{\mu}$ is
written as
\begin{align}
t_{\mu \nu} = \left( \frac{v_{\mu}}{v_{\nu}} \right)^{1/2} T_{\mu \nu},
\label{Trans}
\end{align}
and the reflection coefficient for incoming channel $\nu$ with velocity
$v_{\nu}$, and out-going channel $\mu$ with velocity $v_{\mu}$ as
\begin{align}
r_{\mu \nu} = \left( \frac{v_{\mu}}{v_{\nu}} \right)^{1/2} R_{\mu \nu}.
\end{align}
These formula can also be used to derive the transmission probability for trilayer case. 

\end{document}